\documentclass[prl,twocolumn,floatfix,amsmath,amsfonts]{revtex4} 

\usepackage{amsmath}
\usepackage{graphicx}

\begin{document}


\title{Stable propagation of pulsed beams
\\ in Kerr focusing media with modulated dispersion}

\author{Mar\'{\i}a I. Rodas-Verde}
\affiliation{\'Area de \'Optica, Facultade de Ciencias de
Ourense,\\Universidade de Vigo, As Lagoas s/n, Ourense, ES-32005
Spain.}

\author{Gaspar D. Montesinos}
\affiliation{Departamento de Matem\'aticas, Escuela T\'ecnica
Superior de Ingenieros Industriales, \\Universidad de Castilla-La
Mancha, 13071 Ciudad Real, Spain}

\author{Humberto Michinel}
\affiliation{\'Area de \'Optica, Facultade de Ciencias de Ourense,\\
Universidade de Vigo, As Lagoas s/n, Ourense, ES-32005 Spain.}

\author{V\'{\i}ctor M. P\'erez-Garc\'{\i}a}
\affiliation{Departamento de Matem\'aticas, Escuela T\'ecnica
Superior de Ingenieros Industriales, \\
Universidad de Castilla-La Mancha, 13071 Ciudad Real, Spain}

\date{\today}


\begin{abstract}
We propose the modulation of dispersion to prevent collapse of
planar pulsed beams which propagate in Kerr-type self-focusing
optical media. As a result, we find a new type of two-dimensional
 spatio-temporal solitons stabilized by
dispersion management. We have studied the existence and properties
of these solitary waves both analytically and numerically. We show
that the adequate choice of the modulation parameters optimizes the
stabilization of the pulse.
\end{abstract}




\maketitle


The analysis of the propagation of high power pulsed laser beams is
among the most active fields of study in Nonlinear Optics, where the
main dynamics phenomenon is the dependence of the refractive index
of the materials with the amplitude of light fields. For propagation
in materials showing a linear dependence of the refractive index
with the laser intensity, the mathematical formulation of the beam
dynamics is adequately described by the cubic nonlinear
Schr\"odinger equation (NLSE) \cite{libros}. In this case, the
excitation of optical solitons is one of the most significant
phenomena \cite{solitons}.

Despite the success of the concept of soliton, these structures
mostly arise in 1+1-dimensional configurations. This is mainly due
to the well known {\em collapse} property of the cubic NLSE in
multi-dimensional scenarios. This implies that a two-dimensional
laser beam which propagates in a Kerr-type nonlinear medium, will be
strongly self-focused to a singularity if the power exceeds a
threshold critical value, whereas for lower powers it will spread as
it propagates. Since collapse prevents the stability of
multidimensional ``soliton bullets" in systems ruled by the cubic
NLSE, a great effort has been devoted to search for systems with
stable solitary waves in multidimensional configurations
\cite{bullets}. It has been recently shown that a modulation  of the
nonlinearity along the propagation direction in the optical material
can be used to prevent the collapse of two-dimensional laser beams
\cite{Berge}. The concept has been extended to the case of several
incoherent optical beams \cite{Montesinos04} and to the case of
matter waves \cite{siempre}.

In this paper we use a similar idea to stabilize
against collapse pulsed laser beams propagating in planar
waveguides. This configuration is of great importance, as it is the
common experimental procedure used for exciting spatial optical
solitons. Instead of making a modulation of
the nonlinearity our idea is to act on the chromatic
dispersion term of the NLSE.  Thus, our procedure resembles that used to obtain
dispersion-managed temporal solitons in optical fibers.

We consider the paraxial propagation along $z$ of a pulsed beam of
finite size in time $t$ and spatially confined by a waveguide along
the $y$-axis. Thus, diffraction only acts in the $x$ direction and
is balanced by the self-focusing nonlinearity given by a refractive
index of the form: $n=n_0+n_2|\vec{E}|^2$. The dynamics of the
slowly-varying amplitude of the pulse $\vec{E}$ is described by a
1+2D NLSE of the form:
\begin{equation}
\label{NLSE}
2i\left(\frac{\partial \vec E}{\partial z}
+k'_0\frac{\partial \vec E}{\partial t}\right)
+k''_0\frac{\partial^2 \vec E}{\partial t^2}
+\frac{1}{k_0}\frac{\partial^2 \vec E}{\partial x^2}
+2k_0\frac{n_2}{n_0}|\vec{E}|^2\vec{E}=0,
\end{equation}
where $k_0,k'_0$ and $k''_0$ are respectively the wavenumber in
vacuum, the inverse of the group velocity, and the group velocity
dispersion coefficient.

As we mentioned above, for continuous beams an adequate modulation
of $n_2$ along $z$ prevents collapse of the light distribution and
yields to stabilized two-dimensional solitons. From a mathematical
point ow view an equivalent effect can be achieved by suitably
modifying the second order derivatives of the propagation equation
\cite{Abd}. However, it is physically impossible to change the sign
of diffraction of a continuous beam during its evolution.
Nevertheless, for the case of pulses, the temporal derivative in Eq.
\eqref{NLSE} can be modulated by changing the dispersion parameter
$k_0''$, which can take positive as well as negative values. In this
paper we will show that it is possible to stabilize a pulsed beam
against collapse by modulating {\em only} the dispersion during beam
propagation, without altering diffraction.

\begin{figure}[htb]
\centerline{\includegraphics[width=8.3cm]{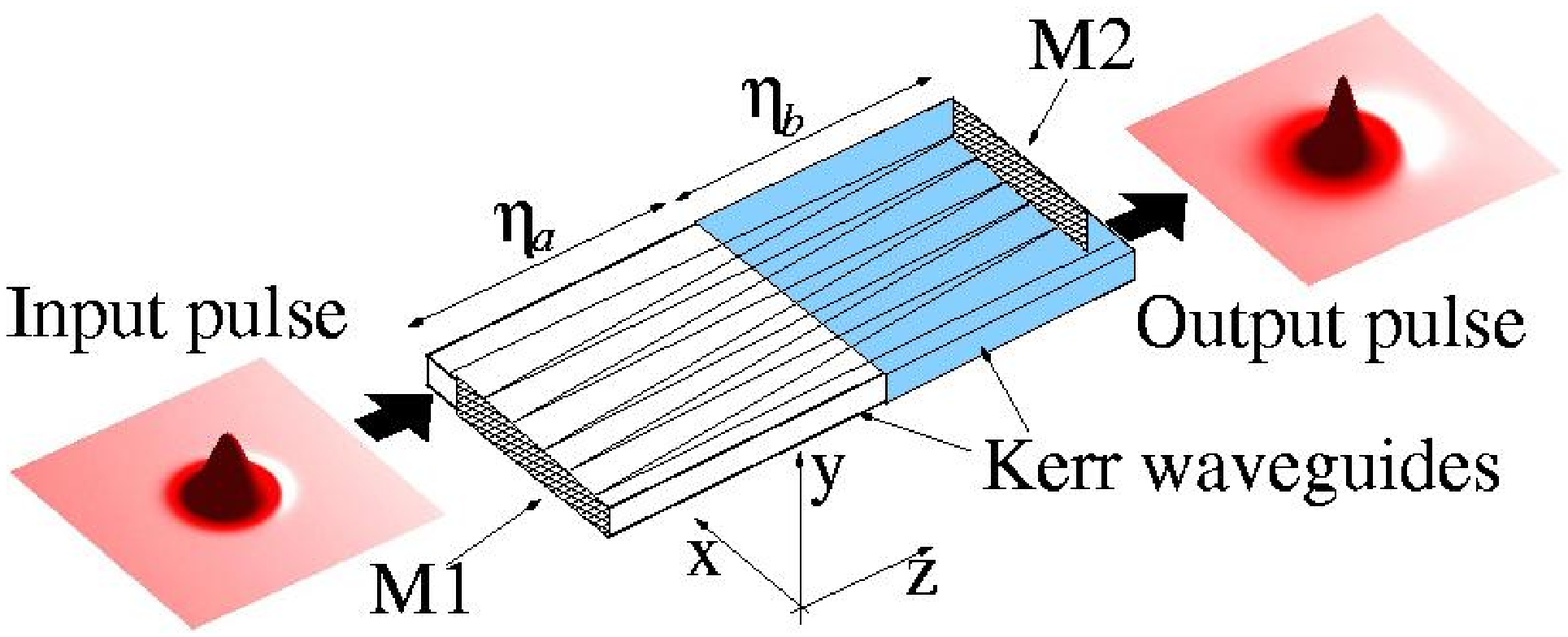}}
\centerline{\includegraphics[width=8.3cm]{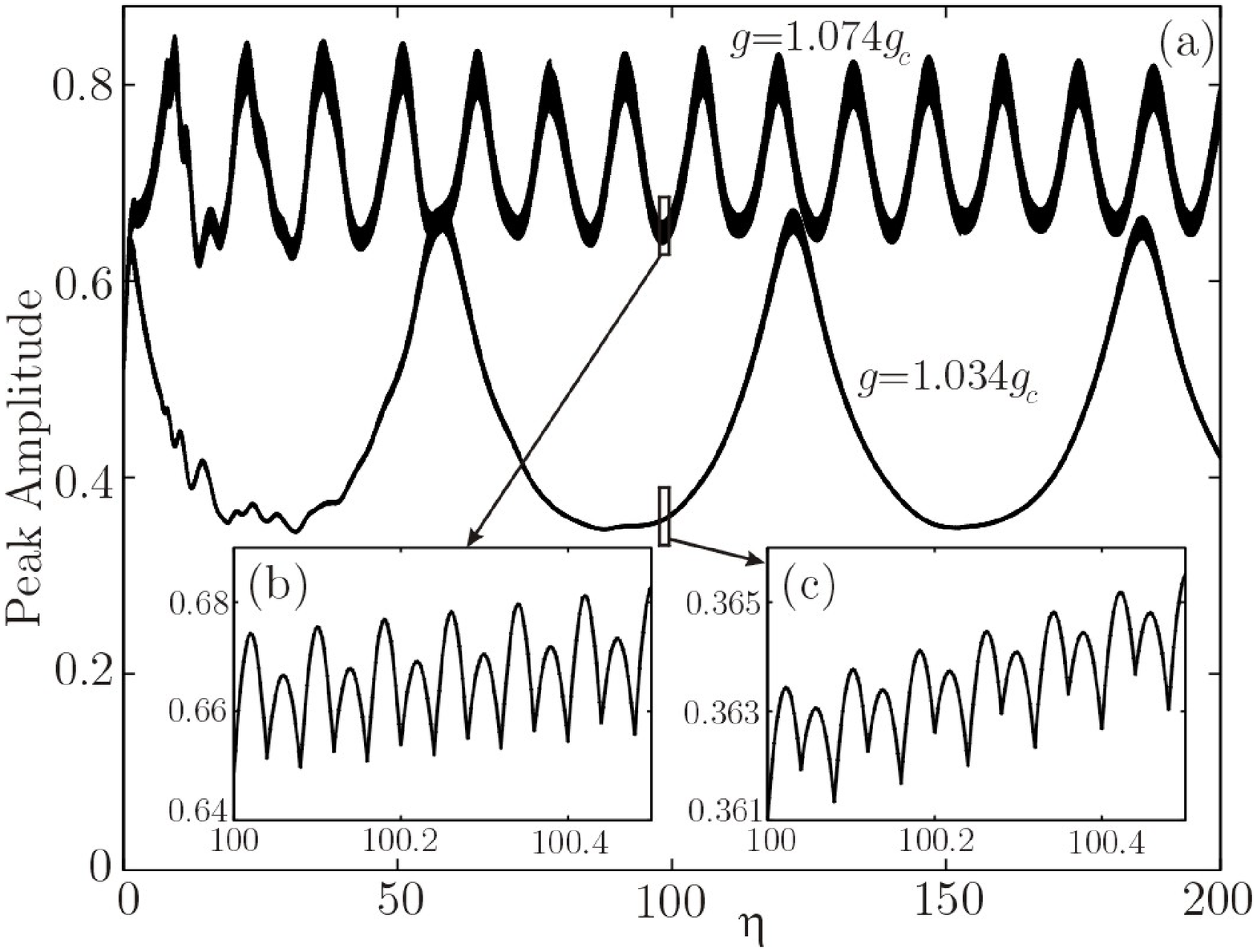}} \caption{
Top: Sketch of the propagation of a pulsed beam through a system
with periodic modulation of dispersion made by joining two different
planar Kerr waveguides. The forward-backward trajectory can be
obtained by reflection at mirrors M1 and M2. Bottom: oscillations of
the peak amplitude of stabilized pulsed beams propagating in the
above system. In (a) the upper curve corresponds to a value
$g=1.074g_c$ (surface plots in top figure). For the lower curve
$g=1.034g_c$. The modulation is of the form: $d_a=1+8$, $d_b=1-8$,
$2\eta_a=2\eta_b=0.04$. In (b) and (c) details of the fast
oscillations are shown.} \label{fig1}
\end{figure}

For a linearly polarized Gaussian pulsed beam with power $P=\int
|E|^2 dz dx$, an initial beam waist $w_{0,x}$ and temporal width
$w_{0,t}$, it is useful to write the above Eq. \eqref{NLSE} in
adimensional form by means of the Fresnel length $F=k_0w_{0,x}^2$ in
the ``reduced time" frame. Thus, we make the rescaling:
$\tau=(t-k'_0z)/w_{0,t}$, $\eta=z/F$, $\chi=x/(F/k_0)^{1/2}$ and
$u=(w_{0,x}w_{0,t}/k_0')^{1/2} E/P^{1/2}$. Then, if we consider that
dispersion is modulated by a periodic function along propagation, we
will finally yield to a generalized NLSE of the following form:
\begin{equation}
\label{NLSE_adim} i \frac{\partial u}{\partial \eta} + \frac{1}{2}
d(\eta)\frac{\partial^2 u}{\partial\tau^2}
+\frac{1}{2}\frac{\partial^2 u}{\partial \chi^2}+g|u|^2u=0.
\end{equation}
Here, $d(\eta)=Fk_0''/w_{0,t}^2$ and $g=k_0 n_2 P F k_0'/n_0 w_{0,x}
w_{0,t}$. It is not obvious what kind of periodic function $g(\eta)$
will be more adequate to stabilize pulsed beams.
A simple choice (and very natural for applications) is to take
$d(\eta)$ as a piecewise constant function of the form $d(\eta)=d_a$
for $\eta\leq 2\eta_a$ and $d(\eta)=d_b$ for $2\eta_a<\eta\leq
2\eta_b$ (see top of Fig. \ref{fig1}) where $d_a$, $d_b$, $\eta_a$
and $\eta_b$ are free parameters that can be varied in order to fit
experimental requirements and to optimize the stabilization of the
pulsed beam.

To get some insight on the dynamics given by Eq. \eqref{NLSE_adim},
we have performed an analytic study by means of the time-dependent
variational approach \cite{Anderson}. Notice that Eq.
\eqref{NLSE_adim} can be obtained from the Lagrangian density:
\begin{equation}
\label{L} 2\mathcal{L}=
i\left(u\dot{u}^{\ast}-u^{\ast}\dot{u}\right)
+d(\eta)\left|\frac{\partial u}{\partial \tau}\right|^{2}
+\left|\frac{\partial u}{\partial
\chi}\right|^{2}-g\left|u\right|^{4},
\end{equation}
where dots denote derivative with respect to $\eta $. We choose a
Gaussian ansatz of the form:
\begin{equation}
\label{trial} u=A\exp\left[-\left(\frac{\chi^{2}}{2w_{\chi}^{2}}
+\frac{\tau^{2}}{2w_{\tau}^{2}}\right)
+i\left(\beta_{\chi}\chi^{2}+\beta_{\tau}\tau^2\right)\right].
\end{equation}
The $\eta-$dependent parameters in the above equation have the
following meaning:  $A$ is the amplitude, $w_{\chi}, w_{\tau}$ are
the spatial and temporal widths and $\beta_{\chi},\beta_{\tau}$ are
the initial curvature and chirp. Although Gaussians are not exact
solutions of Eq. (\ref{NLSE_adim}), our choice simplifies the
calculations and is a usual lithg distribution in experiments. The
standard variational calculations \cite{Anderson} lead to the
equations describing the dynamics of the pulsed beam spatial and
temporal widths:
\begin{subequations}
\label{w}
\begin{eqnarray}
\ddot{w}_{\chi}&=&\frac{1}{w_{\chi}^{3}}
-\frac{1}{2\pi}\frac{g}{w_{\chi}^{2}w_{\tau}},\\
\ddot{w}_{\tau}&=&\frac{d(\eta)}{w_{\tau}^{3}}
-\frac{1}{2\pi}\frac{g}{w_{\tau}^{2}w_{\chi}},
\end{eqnarray}
\end{subequations}
Numerical simulations of Eqs. (\ref{w}) show that the beam can be
stabilized for many choices of the model parameters. These equations
also predict two types of oscillations for width $w_\tau$: a fast
one due to the modulation of dispersion and a slow one which almost
coincides with the variation of $w_\chi$. This low frequency
oscillation is generated by the internal nonlinear dynamics of the
system.

Variational models usually provide only simple and intuitive
predictions that must be complemented with the integration of Eq.
(\ref{NLSE_adim}). All the results to be presented in this paper are
based on direct simulations of Eq. (\ref{NLSE_adim}) using a
split-step Fourier method on a  $520\times 520$ grid and absorbing
boundary conditions to get rid of radiation. In Fig. \ref{fig1} we
plot the evolution of the peak amplitude of two different pulsed
beams in a system sketched in Fig. \ref{fig1} (top). In Fig.
\ref{fig1}(a) the upper curve corresponds to a value $g=1.074g_c$,
being $g_c>0$ the critical value of the nonlinearity for collapse.
For the lower curve $g=1.034g_c$. The modulation is of the form:
$d_a=1+8$, $d_b=1-8$, $2\eta_a=2\eta_b=0.04$. In (b) and (c) details
of the fast oscillations are shown.

\begin{figure}[htb]
\centerline{\includegraphics[width=8.3cm]{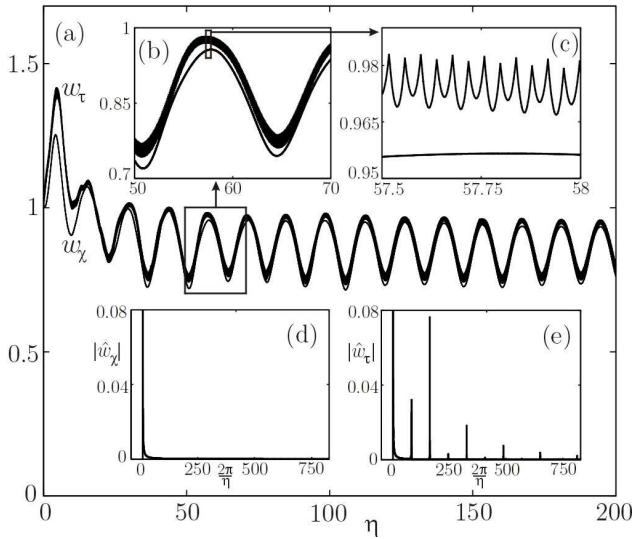}} \caption{(a)
Oscillations of the spatial (lower) and temporal (upper) widths of
the stabilized soliton with $g=1.074g_c$ from Fig. \ref{fig1}(a).
Insets (b) and (c) display details of the curves. Insets (d) and (e)
show the frequency spectra of the spatial and temporal width
oscillations, respectively.} \label{fig2}
\end{figure}

In Fig. \ref{fig2} we plot the width oscillations of the stabilized
soliton from Fig. \ref{fig1} with $g=1.074 g_c$. As the variational
equations predict, the spatial width only displays a low frequency
oscillation, whereas the temporal width presents small amplitude
fast variations over the same low frequency main oscillation. This
is clearly seen in the frequency spectra displayed in Fig.
\ref{fig2}(d) and Fig. \ref{fig2}(e).

\begin{figure}[htb]
\centerline{\includegraphics[width=8.3cm]{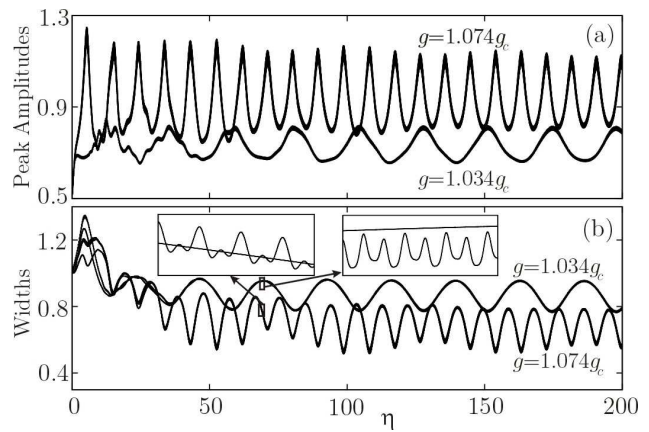}}
\caption{Oscillations of the peak amplitudes (a) and widths (b) of
the same input solitons from Fig. \ref{fig1}. In this case the
modulation is given by $d_a=1+8$, $d_b=1-8$, $2\eta_a=2\eta_b=0.02$.
In (b) the oscillations of the spatial and temporal widths are
overlapped (insets).} \label{fig3}
\end{figure}

In Fig. \ref{fig3} we plot the results of simulation of Eq.
\eqref{NLSE_adim} for the same input pulsed beams of Fig.
\ref{fig1}. Now we take waveguides of half size, resulting in a
frequency of modulation which is twice that of Fig. \ref{fig1}.
Stabilization of both beams is achieved, showing the robustness
of this idea.

As a conclusion we can say that in this paper we have described
two-dimensional spatio-temporal solitons which are
stabilized against collapse by means of dispersion management. We
have studied their stability and properties both analytically and
numerically. We have shown that the adequate choice of the
modulation parameters will optimize the stabilization of the pulse.
Our results are of importance in the field of high power pulse
propagation in nonlinear optical materials.

A remaining open question concerns the stability of
higher-dimensional beams. In fact, at the present time it is not
clear the mechanism of stabilization of fully three dimensional
solitons which have been predicted to collapse in the case of
modulation of nonlinearity \cite{siempre}. Thus, our results on
dispersion management would be an important step forward in
stabilization of light bullets.

G. D. M. and V. M. P-G. are partially supported by Ministerio de
Educaci\'on y Ciencia under grant BFM2003-02832 and Consejer\'{\i}a
de Educaci\'on y Ciencia of the Junta de Comunidades de Castilla-La
Mancha under grant PAC-02-002. G. D. M. acknowledges support from
grant AP2001-0535 from MECD.



\begin{thebibliography}{99}

%
\bibitem{libros}{C. Sulem and P. Sulem, {\em The nonlinear Schr{\"o}dinger equation:
Self-focusing and wave collapse}, (Springer, Berlin, 2000); Yu. S.
Kivshar and G. P. Agrawal, {\em Optical Solitons: From Fibers to
Photonic Crystals}, (Academic Press, San Diego, 2003).}

\bibitem{solitons}{V. E. Zakharov and A. B. Shabat, Sov. Phys. JETP.
\textbf{34}, 62 (1972); A. Hasegawa and F. Tapper, Appl. Phys. Lett.
\textbf{23}, 142-144 (1973); S. Maneuf, A. Barthelemy and C.
Froehly, J. Optics. \textbf{17}, 139-145 (1986).}

%
%
%
%

\bibitem{bullets}{See e.g. F. Wise and P. Di Trapani, Opt. Photonics
News \textbf{13(2)}, 28 (2002); M. Segev and G.I. Stegeman, Phys.
Today {\bf 51}, No. 8, 42 (1998);
 H. Michinel, R. de
la Fuente, J. Linares, Appl. Opt. \textbf{33}, 3384-3390 (1994).}

\bibitem{Berge}{L. Berge, V. K. Mezentsev, J. J. Rasmussen, P. L.
Christiansen and Y. B. Gaididei, Opt. Lett. \textbf{25}, 1037
(2000); I.Towers and B.A.Malomed, J. Opt. Soc. Am. B \textbf{19},
537 (2002).}


\bibitem{Montesinos04}{G. D. Montesinos,  V. M. P\'erez-Garc\'{\i}a and H. Michinel, Phys. Rev. Lett. \textbf{92}, 133901-1 (2004).}

\bibitem{siempre}{H. Saito and M. Ueda, Phys. Rev. Lett. \textbf{90}, 040403 (2003); F. Abdullaev, J. G. Caputo, R. A. Kraenkel, and B. A.
Malomed, Phys. Rev. A \textbf{67}, 013605 (2003); G. D. Montesinos,
V. M. P\'erez-Garc\'{\i}a and P. Torres, Physica D \textbf{191},
193-210 (2004).}



\bibitem{Abd}{F. Adbullaev, B. Baizakov, and M Salerno,
Phys. Rev. E    \textbf{68}, 066605 (2003).}

\bibitem{Anderson}{D. Anderson and M. Lisak, Phys. Rev. A \textbf{32} 2270 (1985).}

%
%
%
%
%

\end{thebibliography}
\end{document}